\documentclass{article}

\usepackage{geometry}
\usepackage{amssymb}    
\usepackage{cancel}
\usepackage[table]{xcolor}
\usepackage{graphicx}
\usepackage{dsfont}

\title{Consistently Orienting Facets in Polygon Meshes by \\Minimizing the Dirichlet Energy of Generalized Winding Numbers}

\author{
\small{Kenshi Takayama}\\
\small{ETH Zurich} \and
\small{Alec Jacobson} \\
\small{ETH Zurich} \and
\small{Ladislav Kavan} \\
\small{University of Pennsylvania} \and
\small{Olga Sorkine-Hornung} \\
\small{ETH Zurich}
}

\date{}

\newcommand{\argmin}{\mathop{\rm arg~min}\limits}

\newcommand{\cheatvspace}[1]{}

\begin{document}
\maketitle
\abstract{

Jacobson et al.~\cite{Jacobson13} hypothesized that the local coherency of the generalized winding number function proposed in that work could be used to correctly determine consistent facet orientations in polygon meshes.
We report on an approach to consistently orienting facets in polygon meshes by minimizing the Dirichlet energy of generalized winding numbers.
While the energy can be concisely formulated and efficiently computed, we found that this approach is fundamentally flawed and is unfortunately not applicable for most handmade meshes shared on popular mesh repositories such as Google 3D Warehouse.

}
\section{Introduction}
The robust inside-outside segmentation method based on the generalized winding number proposed by Jacobson et al.~\cite{Jacobson13} assumes that the polygonal facets in the input mesh are consistently oriented.
If this assumption does not hold, which is often the case for hand-made meshes shared on the Internet (e.g., Google 3D Warehouse), it is necessary to first correct the facet orientations.
In this technical report, we present an approach to correcting inconsistent facet orientations based on the Dirichlet energy of the generalized winding number.
Our motivation came from the fact that if the mesh is watertight and its facets are consistently oriented, the generalized winding number is piecewise constant, and hence its Dirichlet energy is zero.
Our intuition was that for a mesh with various imperfections such as holes and intersections, more consistently oriented facets would result in a smaller Dirichlet energy of the corresponding generalized winding number.
%

\section{Method}

\subsection{Patch extraction}

As a preprocessing step, facets that are topologically connected and forming a two-manifold surface are grouped together as a \emph{patch}, similar to Borodin et al.'s method~\cite{Borodin04}. If the mesh has non-manifold edges, the mesh is split around these edges, and multiple patches are generated.
All facets belonging to the same patch share the same decision about whether they should be flipped or not in the output.
Given $n$ patches, our goal is to seek an assignment for $n$ binary variables representing whether the $i$-th patch $\mathcal{P}_i$ should be flipped or not ($1\leq i\leq n$).

\subsection{Generalized winding numbers}

Let us first review the concept of generalized winding numbers~\cite{Jacobson13}.
The generalized winding number (hereafter we drop the word ``generalized'') at a point $\mathbf{p}\in \mathds{R}^3$ is defined as
\begin{equation}
w(\mathbf{p})=\frac{1}{4\pi}\sum_{\tau\in\mathcal{M}}\theta_\tau(\mathbf{p}) \ ,
\end{equation}
where $\mathcal{M}$ is the input mesh (denoted as a set of facets), $\tau\in\mathcal{M}$ is a facet in $\mathcal{M}$, and $\theta_\tau$ is the signed solid angle subtended by $\tau$ viewed from $\mathbf{p}$. Note that the sign of the angle is defined to be positive for points in the half space corresponding to the back side of a facet. In particular, the winding number of points inside and outside a watertight mesh is 1 and 0, respectively.

\subsection{Dirichlet energy of the winding number}

Since the winding number of a watertight and consistently oriented mesh is piecewise constant and hence its Dirichlet energy is zero, our intuition is that the Dirichlet energy of the winding number for an imperfect mesh should decrease as its facet orientations become more consistent.
We give a mathematical formulation of the Dirichlet energy in the following.

For each $i$-th patch $\mathcal{P}_i \subseteq \mathcal{M} \ (1 \leq i \leq n)$, we want to determine a binary variable $s_i\in\{+1, -1\}$ called \emph{sign}, where $s_i=-1$ means that all facets in $\mathcal{P}_i$ should be flipped in the output.
Now we can define the winding number with respect to the given sign $\mathbf{s}=(s_1, \cdots, s_n)$ as
\begin{equation}
w_{\mathbf{s}}(\mathbf{p})=\frac{1}{4\pi}\sum_{i=1}^n s_i \sum_{\tau\in\mathcal{P}_i}\theta_\tau(\mathbf{p}) \ .
\end{equation}
Note that the orientation of each triangle $\tau$ is fixed to the original configuration when computing $\theta_\tau$.
The Dirichlet energy of $w_{\mathbf{s}}$ denoted as $E(\mathbf{s})$ is a function of $\mathbf{s}$ and is defined as the integral of the gradient magnitude squared of $w_{\mathbf{s}}$ over the volumetric domain $\Omega=\mathds{R}^3\setminus\mathcal{M}$:
\begin{equation}
E(\mathbf{s})=\int_{\mathbf{p}\in\Omega}\left\| \nabla w_\mathbf{s}(\mathbf{p})\right\|^2dV \ .
\end{equation}
Reorganizing the integrand as
\begin{eqnarray}
\left\| \nabla w_\mathbf{s}(\mathbf{p})\right\|^2
&=& \left\|\frac{1}{4\pi} \sum_{i=1}^n s_i \sum_{\tau\in\mathcal{P}_i} \nabla \theta_\tau(\mathbf{p}) \right\|^2    \\
&=& \frac{1}{16\pi^2}\left(  \sum_{i=1}^n s_i \sum_{\tau\in\mathcal{P}_i} \nabla \theta_\tau(\mathbf{p}) \right) \cdot \left( \sum_{j=1}^n s_j \sum_{\mu\in\mathcal{P}_j} \nabla \theta_\mu(\mathbf{p}) \right)    \\
&=& \frac{1}{16\pi^2}\sum_{i=1}^n \sum_{j=1}^n s_i s_j \left( \sum_{\tau\in\mathcal{P}_i} \nabla \theta_\tau(\mathbf{p}) \right) \cdot \left( \sum_{\mu\in\mathcal{P}_j} \nabla \theta_\mu(\mathbf{p}) \right),
\end{eqnarray}
the energy can be written concisely in quadratic form as
\begin{equation}
E(\mathbf{s})=\mathbf{s}^\top\mathbf{Q} \ \mathbf{s} \ ,
\end{equation}
where $Q$ is an $n$-by-$n$ matrix with elements
\begin{equation}
\label{eqn:qij}
Q_{ij}=\frac{1}{16\pi^2}\sum_{\tau\in\mathcal{P}_i}\sum_{\mu\in \mathcal{P}_j}\int_{\mathbf{p}\in\Omega}\nabla\theta_\tau(\mathbf{p})\cdot\nabla\theta_\mu(\mathbf{p}) \ dV \ .
\end{equation}
$Q$ is fixed per model independently of $\mathbf{s}$. 
Our goal is to find a set of facet orientations $\mathbf{s}^*$ that gives the smallest Dirichlet energy:
\begin{equation}
\label{eqn:goal}
\mathbf{s}^* = \argmin_{\mathbf{s}\in\{-1,+1\}^n} E(\mathbf{s}).
\end{equation}
Since $s_i^2 =1$, the energy can be further reorganized as
\begin{equation}
E(\mathbf{s}) = 2\sum_{i=1}^{n-1}\sum_{j=i+1}^n s_i s_j Q_{ij} + \underbrace{\sum_{i=1}^n Q_{ii}}_{\mathrm{const.}} \ .
\end{equation}
Ignoring the diagonals of $\mathbf{Q}$, we can also consider
\begin{equation}
\label{eqn:energy-no-diag}
\tilde{E}(\mathbf{s}) = \sum_{i=1}^{n-1}\sum_{j=i+1}^n s_i s_j Q_{ij}
\end{equation}
as the energy instead.

\subsection{Computing the Dirichlet energy}

We need to compute $Q_{ij}$ in order to evaluate $E(\mathbf{s})$. 
A straightforward approach would be to tessellate the finite volume around the input mesh with sufficiently large volume and fine resolution to capture the variations of the winding numbers and then numerically compute the integral over this tessellation.
However, this approach can be computationally expensive, especially when the mesh contains many small
details and a fine tessellation is needed for achieving sufficient accuracy.

Fortunately, since the winding number function is harmonic (i.e., $\Delta w(\mathbf{p})=0$)~\cite{Jacobson13}, we can convert the volumetric integral over the domain $\Omega$ into a surface integral over the domain boundary $\partial\Omega$ using Green's first identity. For a once differentiable scalar function $\psi$ and a twice differentiable scalar function $\phi$, Green's first identity states
\begin{equation}
\int_{\Omega}\psi\Delta\phi \ dV + \int_{\Omega}\nabla\psi\cdot\nabla\phi \ dV=\int_{\partial\Omega}\psi\left(\nabla\phi\cdot\mathbf{n}\right)dS
\end{equation}
where $\mathbf{n}$ is an outgoing normal at the domain boundary (the positional
variable $\mathbf{p}$ is omitted here for brevity). By substituting
$\psi=\theta_\tau$ and $\phi=\theta_\mu$ and recalling that winding numbers are harmonic, we get
\begin{equation}
\cancelto{0}{\int_{\Omega}\theta_\tau\Delta\theta_\mu \ dV} + \int_{\Omega}\nabla\theta_\tau\cdot\nabla\theta_\mu \ dV = \int_{\partial\Omega}\theta_\tau(\nabla\theta_\mu\cdot\mathbf{n})dS \ ,
\end{equation}
which shows that the volumetric integral in (\ref{eqn:qij}) is now converted to a surface integral.

\begin{figure}[h]
\centering
\includegraphics[height=5cm]{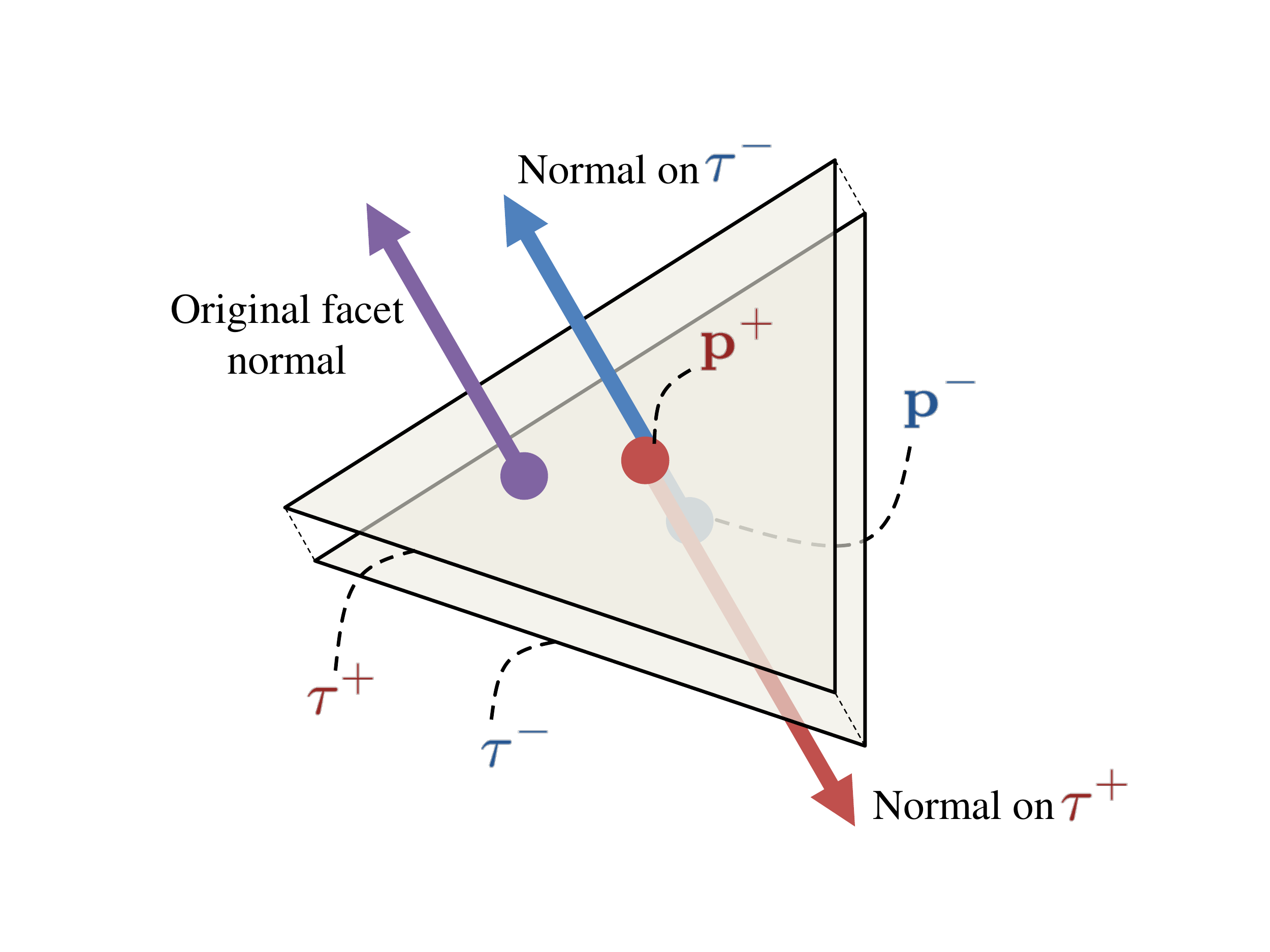}
\caption{Notations for the domain boundary.}
\label{fig-domain}
\end{figure}

To compute this surface integral, we make the following notes on the domain and its boundary. The domain $\Omega$ consists of the entire 3D space $\mathds{R}^3$ with a hole of infinitesimally thin slab carved out by each facet $\tau\in\mathcal{M}$, and the domain boundary $\partial\Omega$ consists of such holes.
Each hole corresponding to $\tau$ has its front and back sides denoted by $\tau^+$ and $\tau^-$, respectively. Note that the domain's outgoing normals on $\tau^+$ and $\tau^-$ are pointing toward the hole and are opposite to each other, see Figure~\ref{fig-domain}. Now the domain boundary can be written as
\begin{equation}
\partial\Omega=\bigcup_{\tau\in\mathcal{M}}\tau^+\cup \tau^-.
\end{equation}
For a facet $\tau\in\mathcal{M}$ and a point on it $\mathbf{p}\in\tau$, let us denote the corresponding point on the front and back sides as $\mathbf{p}^+\in \tau^+$ and $\mathbf{p}^-\in \tau^-$, respectively. Then, it is easy to see that the following properties hold:
\begin{eqnarray}
\theta_\tau(\mathbf{p}^+) &=& -\theta_\tau(\mathbf{p}^-) = -2\pi \\
\nabla\theta_\tau(\mathbf{p}^+) &=& \nabla\theta_\tau(\mathbf{p}^-) \ ,
\end{eqnarray}
where the latter is due to the derivative of the arctangent.
Note that $\theta_\tau$ and $\nabla\theta_\tau$ are undefined along the edges of $\tau$;
we simply ignore the edge parts (i.e., parts connecting $\tau^+$ and $\tau^-$) in the surface integral, since the measure of all the edges vanishes (as opposed to the measure of the facets' interiors).

For any other third facet $\gamma\neq\tau$, $\theta_\tau$ and $\nabla\theta_\mu$ take the same value on $\gamma^+$ and $\gamma^-$. Therefore,
\begin{eqnarray}
\int_{\gamma^+ \cup \gamma^-}\theta_\tau(\nabla\theta_\mu\cdot\mathbf{n})dS 
&=& \int_{\gamma^+} \theta_\tau(\nabla\theta_\mu\cdot\mathbf{n})dS +
    \int_{\gamma^-} \theta_\tau(\nabla\theta_\mu\cdot\mathbf{n})dS    \\
\label{eqn:18}
&=& \int_{\gamma^+} \theta_\tau(\nabla\theta_\mu\cdot\mathbf{n})dS +  
    \int_{\gamma^+} \theta_\tau(\nabla\theta_\mu\cdot(-\mathbf{n}))dS \\
&=& 0
\end{eqnarray}
where in (\ref{eqn:18}) we replaced the integral over $\gamma^-$ with an integral over $\gamma^+$ with a negated surface normal. Therefore, the integral over $\partial\Omega$ is reduced to an integral over $\tau^+\cup\tau^-$, i.e.,
\begin{eqnarray}
\int_{\partial\Omega}\theta_\tau(\nabla\theta_\mu\cdot\mathbf{n})dS &=& \sum_{\gamma\in\mathcal{M}}\int_{\gamma^+ \cup \gamma^-}\theta_\tau(\nabla\theta_\mu\cdot\mathbf{n})dS \\
&=& \int_{\tau^+ \cup \tau^-}\theta_\tau(\nabla\theta_\mu\cdot\mathbf{n})dS \ .
\end{eqnarray}
Using the properties above, we arrive at:
\begin{eqnarray}
\int_{\tau^+ \cup \tau^-}\theta_\tau(\nabla\theta_\mu\cdot\mathbf{n})dS 
&=& \int_{\tau^+}(-2\pi)(\nabla\theta_\mu\cdot\mathbf{n})dS +        
    \int_{\tau^-}( 2\pi)(\nabla\theta_\mu\cdot\mathbf{n})dS          \\
&=& -2\pi\int_{\tau^+}\nabla\theta_\mu\cdot  \mathbf{n}  \ dS + 
     2\pi\int_{\tau^+}\nabla\theta_\mu\cdot(-\mathbf{n}) \ dS   \\
&=& -4\pi\int_{\tau^+}\nabla\theta_\mu\cdot\mathbf{n} \ dS \ .
\end{eqnarray}
In summary, $Q_{ij}$ is computed as
\begin{equation}
Q_{ij} = \sum_{\tau\in\mathcal{P}_i}\sum_{\mu\in\mathcal{P}_j} F(\tau, \mu)
\end{equation}
where
\begin{equation}
F(\tau, \mu) = -\frac{1}{4\pi}\int_{\mathbf{p} \in \tau^+} \nabla\theta_\mu(\mathbf{p})\cdot\mathbf{n}(\mathbf{p}) \ dS \ .
\end{equation}

\subsection{Convergence of the Dirichlet energy}

By deriving an analytic form of $F(\tau, \mu)$ using \textsc{Maple}, we found that it evaluates to infinity (either positive or negative depending on relative orientation of $\tau$ and $\mu$) when some edges of $\tau$ and $\mu$ coincide geometrically. 
An intuitive explanation for this is that the magnitude of $\nabla\theta_\mu(\mathbf{p})$ becomes infinitely large when $\mathbf{p} \in \tau^+$ goes infinitesimally close to the edges of $\tau$ that are geometrically coinciding with the edges of $\mu$.
In particular, the diagonal entries $Q_{ii}$ always evaluate to positive infinity.
Although we can ignore the diagonals using (\ref{eqn:energy-no-diag}), off-diagonal entries $Q_{ij}$ can still evaluate to infinity if some boundary edges of $\mathcal{P}_i$ and $\mathcal{P}_j$ coincide geometrically.
In our experiments, we numerically evaluate the energy with a sufficiently fine discretization of $\partial\Omega$ where the finite entries in $\mathbf{Q}$ are converged.



\subsection{Finding the minimizer of the Dirichlet energy}

The minimization problem in (\ref{eqn:goal}) is referred to as Quadratic Unconstrained Binary Optimization (QUBO).
QUBO is NP-hard, and various heuristic methods exist~\cite{Boros07}.
In our experiments, we take a brute force approach and exhaustively search for the optimal solution, so that we can examine the nature of our approach with small toy examples.

\section{Results}

\begin{figure}[h]
\centering
\includegraphics[height=4.5cm]{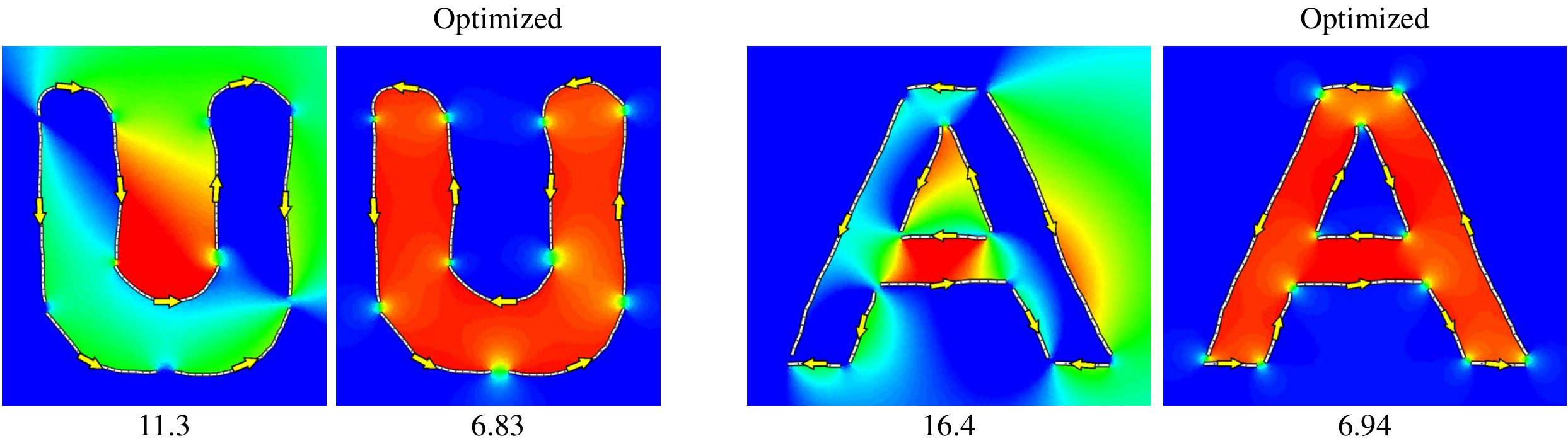}
\caption{Examples where our approach produces reasonable results.
The input polylines are represented by white lines with their orientations indicated by yellow arrows. 
The number below each image is the Dirichlet energy of the corresponding winding number function.
The color maps represent the associated winding numbers, where higher winding numbers correspond to warmer colors (the actual mapping is slightly adjusted for each image to better visualize the distribution).}
\label{fig:good}
\end{figure}

\begin{figure}[h]
\centering
\includegraphics[height=4.5cm]{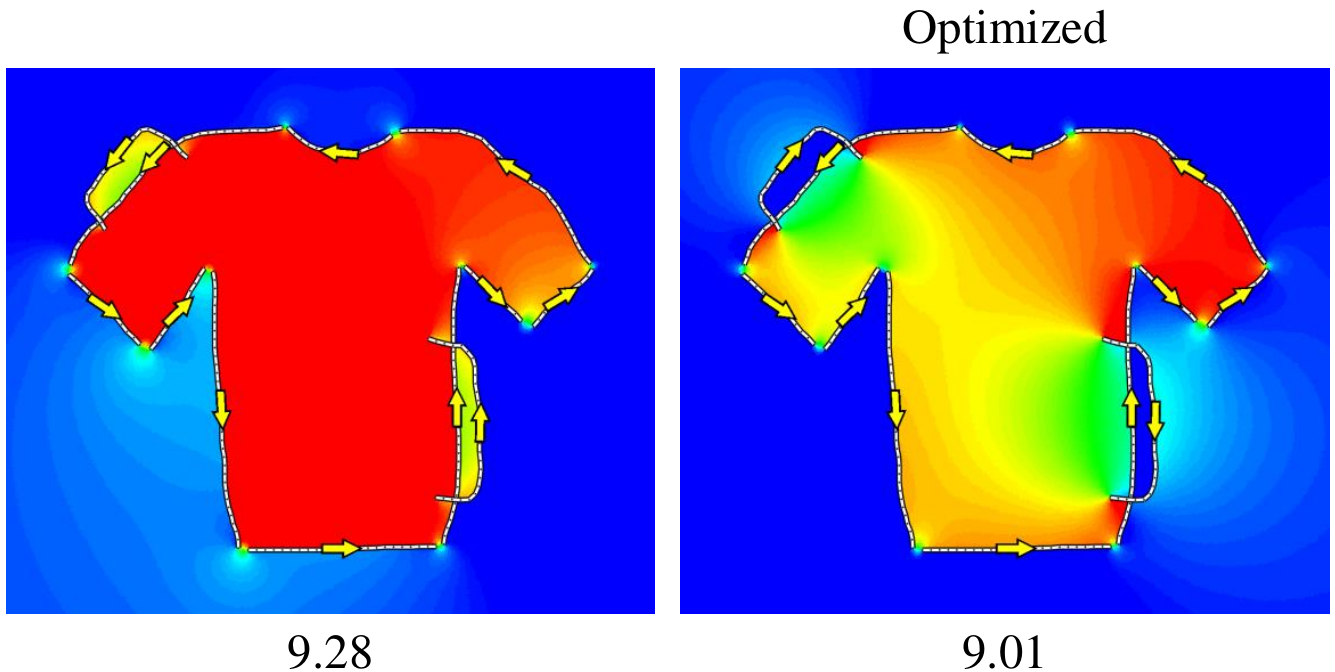}
\caption{An example representing a torso with two small objects attached to the shoulder and the side. Our approach results in those small attachments oriented inside-out.}
\label{fig:bad}
\end{figure}

We implemented our approach in 3D as well as in 2D where the input consists of open or closed polylines (still referred to as ``patches'' in the following) and the same formulation applies.
Although here we show our experiments only in 2D for the ease of visualization, the same argument also holds in 3D.
Figure~\ref{fig:good} shows cases where the input corresponds to the boundary of a solid shape with some missing pieces. In such cases, our algorithm produces the expected results, i.e., it orients the facets consistently and correctly identifies the shape's interior.
On the other hand, Figure~\ref{fig:bad} shows a case where our approach is fundamentally flawed: the input has some small patches attached to larger patches, and those small patches are oriented inside-out.
The fundamental reason for this is that the Dirichlet energy only measures the variation of the function while ignoring the sign, and hence there is no penalty for orienting facets inside-out.

\begin{figure}[h]
\centering
\includegraphics[width=0.7\hsize]{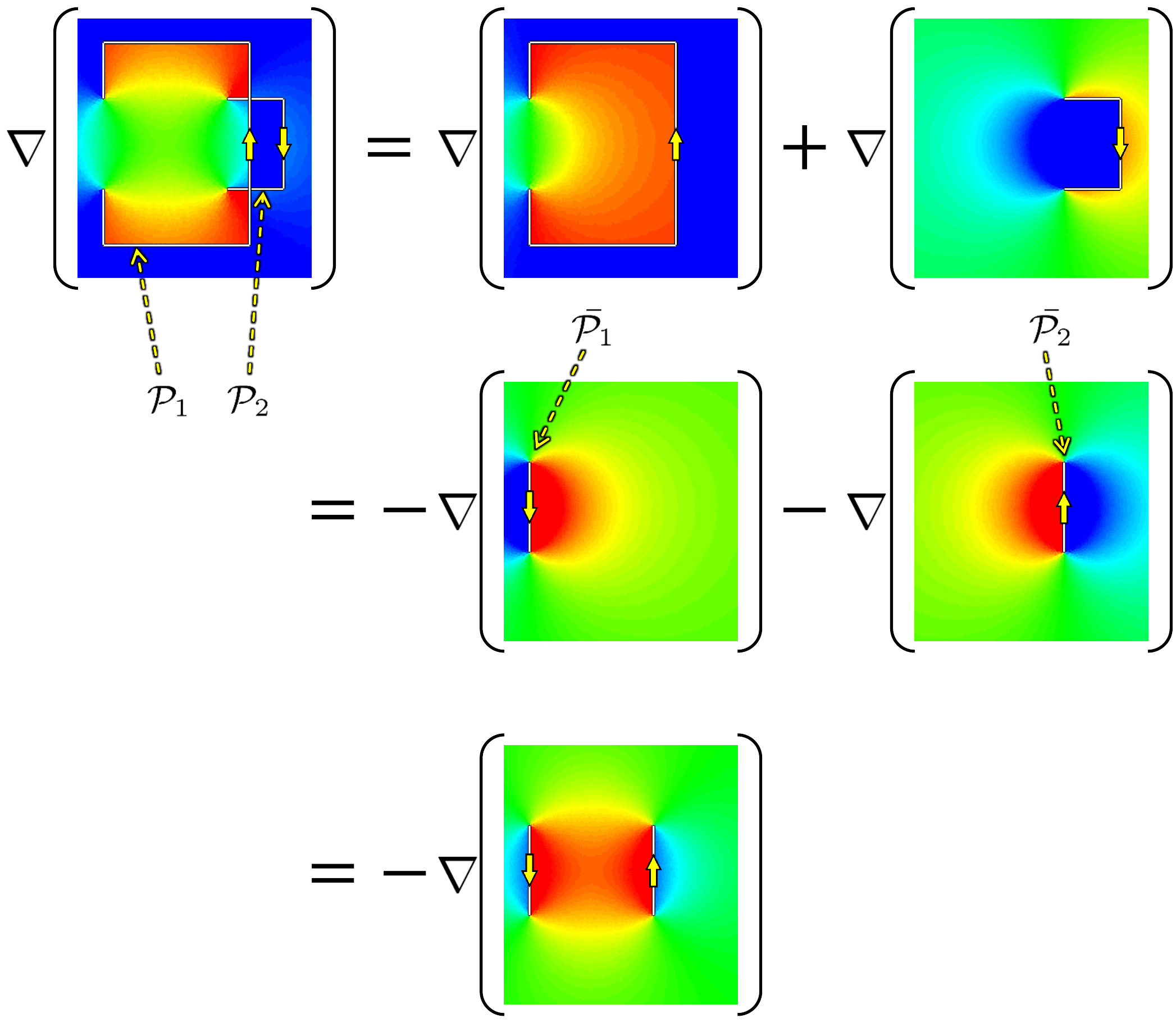}
\caption{An example showing the reason of why small attachments are oriented inside-out in our approach. Note again that the mapping between the winding number and the color is adjusted in each image to better visualize the function.}
\label{fig-reason}
\end{figure}

For an intuitive explanation of this behavior, it is useful to consider the \emph{closure} of a patch $\mathcal{P}$ denoted as $\bar{\mathcal{P}}$~\cite{Jacobson13}, where the union of $\mathcal{P}$ and $\bar{\mathcal{P}}$ bounds a finite volumetric domain $\Omega_\mathcal{P}$ (i.e., $\partial\Omega_\mathcal{P}=\mathcal{P}\cup\bar{\mathcal{P}}$).
Denoting the winding number defined by a patch $\mathcal{P}$ as $w_\mathcal{P}(\mathbf{p})$, the following holds:
\begin{eqnarray}
\label{eqn-closure}
w_\mathcal{P}(\mathbf{p}) + w_{\bar{\mathcal{P}}}(\mathbf{p}) =
\left\{
\begin{array}{ll}
1, & \mathbf{p}\in\Omega_\mathcal{P} \\
0, & \mathrm{otherwise}
\end{array}
\right. .
\end{eqnarray}
Since the right hand side of (\ref{eqn-closure}) is piecewise-constant, the following also holds:
\begin{equation}
\label{eqn-closure-gradient}
\nabla w_\mathcal{P}(\mathbf{p})=-\nabla w_{\bar{\mathcal{P}}}(\mathbf{p}) \ .
\end{equation}
Let us now consider a simple case with two patches $\mathcal{P}_1$ and $\mathcal{P}_2$ as shown in Figure~\ref{fig-reason}.
We can decompose the gradient and apply the above observation as:
\begin{eqnarray}
\nabla w_{\mathcal{P}_1\cup\mathcal{P}_2}(\mathbf{p})
&=& \nabla w_{\mathcal{P}_1}(\mathbf{p}) + \nabla w_{\mathcal{P}_2}(\mathbf{p}) \\
&=& -\nabla w_{\bar{\mathcal{P}_1}}(\mathbf{p}) - \nabla w_{\bar{\mathcal{P}_2}}(\mathbf{p}) \ .
\end{eqnarray}
It is clear in the figure that the current configuration of $\bar{\mathcal{P}_1}$ and $\bar{\mathcal{P}_2}$ yields smaller Dirichlet energy than the other configuration, and hence $\mathcal{P}_2$ is oriented this way in our approach.

\section{Conclusion and future work}

We conclude that the presented approach has a fundamental flaw and is not applicable for most handmade meshes.
One direction for further exploration is to apply a similar approach to surface reconstruction from an unoriented point cloud~\cite{Mullen10,Liu10}, where only the positions of the points are given and finding good normals for the points is critical.
In this case, the unknowns are the continuous normal directions, potentially making the problem easier.
A key question, however, would be how to generalize the definition of winding numbers to point clouds.

\section*{Acknowledgements}

We are grateful to Peter Kaufmann, Wenzel Jakob, Nobuyuki Umetani, Melina Skouras, Ilya Baran, and Ryan Schmidt for their advice and feedback.
This work was supported in part by the ERC grant iModel (StG-2012-306877), by an SNF award 200021\_137879, by an NSF grant IIS-1350330 and by a gift from Adobe Research.
Kenshi Takayama's work was funded by JSPS Postdoctoral Fellowships for Research Abroad.
Alec Jacobson's work was supported by an Intel Doctoral Fellowship.

\bibliographystyle{alpha}
\bibliography{references}

\end{document}